# Symmetric Collaborative Filtering Using the Noisy Sensor Model

Rita Sharma
Department of Computer Science
University of British Columbia
Vancouver, BC V6T 1Z4
rsharma@cs.ubc.ca

David Poole
Department of Computer Science
University of British Columbia
Vancouver, BC V6T 1Z4
poole@cs.ubc.ca

## Abstract

Collaborative filtering is the process of making recommendations regarding the potential preference of a user, for example shopping on the Internet, based on the preference ratings of the user and a number of other users for various items. This paper considers collaborative filtering based on explicit multi-valued ratings. To evaluate the algorithms, we consider only *pure* collaborative filtering, using ratings exclusively, and no other information about the people or items.

Our approach is to predict a user's preferences regarding a particular item by using other people who rated that item and other items rated by the user as noisy sensors. The noisy sensor model uses Bayes' theorem to compute the probability distribution for the user's rating of a new item. We give two variant models: in one, we learn a *classical normal linear regression* model of how users rate items; in another, we assume different users rate items the same, but the accuracy of the sensors needs to be learned. We compare these variant models with state-of-the-art techniques and show how they are significantly better, whether a user has rated only two items or many. We report empirical results using the EachMovie database [1] of movie ratings. We also show that by considering items similarity along with the users similarity, the accuracy of the prediction increases.

## 1  Introduction

Collaborative filtering is a key technology used to build Recommender Systems on the Internet. It has been used by Recommender Systems such as Amazon.com — a book store on the web, CDNow.com — a CD store on the web, and MovieFinder.com — a movie site on the internet [Schafer, Konstan and Riedl, 1999].

Collaborative filtering (CF) is the process of making predictions whether a user will prefer a particular item, given his or her ratings on other items and given other people's ratings on various items including the one in question. CF relies on the fact that people's preferences are not randomly distributed; there are patterns within the preferences of a person and among similar groups of people, creating correlation. The user for whom we are predicting a rating is called the active user. In collaborative filtering, the main premise is that the active user will prefer items which likeminded people prefer, or even that dissimilar people don't prefer. The problem can be formalized: given a set of ratings for various user-item pairs, predict a rating for a new user-item pair. It is interesting that the abstract problem is symmetric between users and items.

Collaborative filtering has been an active area of research in recent years. Several collaborative filtering algorithms have been suggested, ranging from binary to non-binary rating, implicit and explicit rating. Initial collaborative filtering algorithms were based on statistical methods using correlation between user preferences [Resnick, Iacovou, Suchak, Bergstrom and Riedl, 1994; Shardanand and Maes, 1995]. These correlation based algorithms predict the active user ratings as a similarity-weighted sum of the other users ratings. These algorithms are also referred to as memory based algorithms [Breese, Heckerman and Kadie, 1998]. Collaborative filtering is different to the standard supervised learning task because there are only two attributes, each with a large domain; it is the structure within the domains that are important to the prediction, but this structure is not provided explicitly. Recently, some researchers have used machine learning methods [Breese et al., 1998; Ungar and Fos-

---

[1] http://research.compaq.com/SRC/eachmovie/

ter, 1998] for collaborative filtering algorithms. These methods essentially discover the hidden attributes for users and items, which explain the similarity between users and items.

Breese et al. [Breese et al., 1998] proposed and evaluated two probabilistic models for model based collaborative filtering algorithms: *cluster models* and *Bayesian networks*. In the cluster model, users with similar preferences are clustered together into classes. The model's parameters, the number of clusters, and the conditional probability of ratings given a class are estimated from the training data. In the Bayesian network, nodes correspond to items in the database. The training data is used to learn the network structure and the conditional probabilities.

Pennock et al.[Pennock, Horvitz, Lawrence and Giles, 2000] proposed a collaborative filtering algorithm called personality diagnosis ($PD$) and showed that $PD$ makes better predictions than other memory and model based algorithms. This algorithm is based on a probabilistic model of how people rate items, which is similar to our noisy sensor model approach.

In this paper we propose and evaluate a probabilistic approach based on a noisy sensor model, which is symmetric between users and items. Our approach is based on the idea that to predict an active user's rating for a particular item, we can use all those people who rated that item and other items rated by the active user as the noisy sensors. We view the noisy sensor model as a belief network. The conditional probability table associated with each sensor reflects the noise in the sensor.

To model how another user (user $u$) can act as a noisy sensor for the active user $a$'s rating, we need to find a relationship between their preferences. Unfortunately, there is usually very little data, so we need to make *a priori* some assumptions about the relationship. Here we give two variants of the general idea for learning the noisy sensor model for explicit multi-valued rating data: one, where we learn a classical normal linear regression model of how users rate items (*Noisy1*); and another, where we assume that the different users rate items the same and learn the accuracy of the sensor(*Noisy2*).

In order to avoid a perfect fit with sparse data we add some dummy points before fitting the relationship. We use *hierarchical prior* to distribute the effect of dummy points over all possible rating pairs.

After learning the noisy sensor model (i.e. the conditional probability table associated with each sensor node), we use Bayes' theorem to compute the probability distribution of the user $a$'s rating of a new item.

We evaluate both *Noisy1* and *Noisy2* on the EachMovie database of movie ratings and compare them to the state-of-the-art techniques. We also show that symmetric collaborative filtering, which employs both user and item similarity, offers better accuracy than asymmetric collaborative filtering.

## 2 Filtering Problem and Mathematical Notation

Let $N$ be the number of users and $M$ be the total number of items in the database. $S$ is an $N \times M$ matrix of all user's ratings for all items; $S_{ui}$ is the rating given by user $u$ to item $i$. Let the ratings be on a cardinal scale with $m$ values that we denote $v_1, v_2, \ldots, v_m$. Then each rating $S_{ui}$ has a domain of possible values $(v_1, v_2, \ldots, v_m)$. In collaborative filtering, $S$, the user-item matrix, is generally very sparse since each user will only have rated a small percentage of the total number of items. Under this formulation, the collaborative filtering problem becomes predicting those $S_{ui}$ which are not defined in $S$, the user-item matrix.

## 3 Collaborative Filtering Using the Noisy Sensor Model

We propose a simple probabilistic approach for symmetric collaborative filtering using the noisy sensor model for predicting the rating by user $a$ (active user) of an item $j$. We use as noisy sensors:

- all users who rated the item $j$
- all items rated by user $a$

The sensor model is depicted as a naive Bayesian network in Figure 1. The direction of the arrow shows that the prediction of user $a$ for item $j$ causes the sensor $u$ to take on a particular prediction for item $j$. The sensor model is the conditional probability table associated with each sensor node. The noise in the sensor is reflected by the probability of incorrect prediction; that is, by the conditional probability table associated with it. To keep the model simple we use the independence assumption that the prediction of any sensor for item $j$ is independent of others, given the prediction of user $a$ for item $j$.

We need the following probabilities for Figure 1:

$Pr(S_{uj}|S_{aj})$ : the probability of user $u$'s prediction for item $j$, given the prediction of user $a$ for item $j$.

$Pr(S_{ak}|S_{aj})$ : the probability of user $a$'s prediction for item $k$, given the prediction of user $a$ for item $j$.

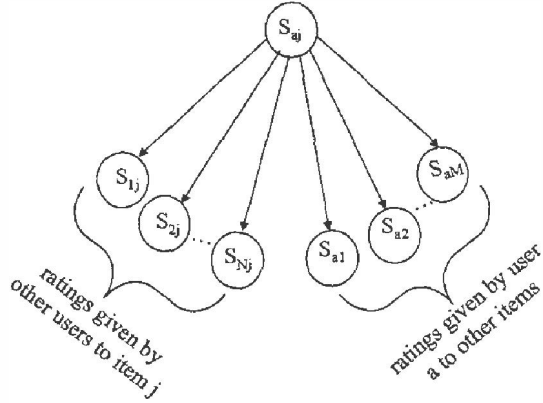

Figure 1: Naive Bayesian Network Semantics for the Noisy Sensor Model

$Pr(S_{aj})$ : the prior probability of active user $a$'s prediction for item $j$.

We compute the prior probability distribution $Pr(S_{aj} = v_i)$ of user $a$'s rating for item $j$ by the fraction of rating $v_i$ in the training data set, where $v_i \in (v_1, v_2, \ldots, v_m)$.

Given the conditional probabilities table for all sensors, we can compute the probability distribution for user $a$'s rating of an unseen item $j$, using the noisy sensor model as described in Figure 1. By applying Bayes' rule we can have the following:

$Pr(S_{aj} | (S_{1j}, \ldots, S_{Nj}) \wedge (S_{a1}, \ldots, S_{aM}))$

$$\propto Pr(S_{aj}) \cdot \prod_{u=1}^{N} Pr(S_{uj}|S_{aj}) \cdot \prod_{k=1}^{M} Pr(S_{ak}|S_{aj}) \quad (1)$$

To use the noisy sensor model for collaborative filtering we need the probability table for probabilities: $Pr(S_{uj}|S_{aj})$ and $Pr(S_{ak}|S_{aj})$.

Consider first the problem of estimating $Pr(S_{uj}|S_{aj})$, which is the problem of estimating user $u$'s rating for item $j$ given user $a$'s rating of it. There is typically sparse data for the $m \times m$ probability table and we need to make some prior assumptions about the relationship. We assume that there is a linear relationship with Gaussian error between the preferences of users and, similarly, between the ratings received by the items. Suppose the rating of user $a$ (the independent variable) is denoted by $x$, and that the rating of user $u$ (the dependent variable) is denoted by $y$. Suppose that user $a$ and user $u$ co-rated $n$ items and their ratings over $n$ co-rated items are denoted by $n$ pairs of observations $(x_1, y_1), (x_2, y_2), \ldots, (x_n, y_n)$. We want to find a straight line which best fits these points. We assume that the mean of $y$ can be expressed as a linear function of independent variable $x$. Since a model based on an independent variable cannot in general predict exactly the observed values of $y$, it is necessary to introduce the error $e_i$. For the $i^{th}$ observation, we have the following:

$y_i = \alpha + \beta x_i + e_i$

We assume that unobserved errors $(e_i)$ are independent and normally distributed with a mean of zero and the variance $\sigma^2$. If $y_i$ is the linear function of $e_i$, which is normally distributed, then $y_i$ is also normally distributed. We assume the same variance for all the observations. The mean and variance of $y_i$ are given thus:

$E(y_i) = \alpha + \beta x_i$

$var(y_i) = \sigma^2$

For the $i^{th}$ observation, the probability distribution function of $y$ which is normally distributed can be written thus:

$$P(y = y_i | x = x_i) = \frac{1}{\sqrt{2\pi\sigma^2}} exp\left[\frac{-1}{2\sigma^2}(y_i - \mu_i)^2\right]$$

where $\mu_i = \alpha + \beta x_i$

The joint probability distribution function (or the *likelihood function* denoted by $LF(\alpha, \beta, \sigma^2)$) is the product of the individual $P(y_i|x_i)$ over all observations.

$$LF(\alpha, \beta, \sigma^2) = \prod_{i=1}^{n} P(y_i|x_i)$$

$$= \frac{1}{\left(\sqrt{2\pi\sigma^2}\right)^n} exp\left[\frac{-1}{2\sigma^2} \sum_{i=1}^{n}(y_i - \mu_i)^2\right]$$

We apply the maximum likelihood method [Gujarati, 1995] to estimate unknown parameters $(\alpha, \beta, \sigma^2)$. The likelihood is maximum at the following values of the parameters:

$\alpha = 1/n \left(\sum y_i - \beta \sum x_i\right)$

$\beta = \frac{n \sum y_i - \sum y_i x_i}{n \sum x_i - \sum x_i^2}$

$\sigma^2 = \frac{1}{n} \sum (y_i - \alpha - \beta x_i)^2 \quad = \frac{1}{n} \sum e_i^2$

After calculating the parameters $\alpha$, $\beta$ and $\sigma^2$ we can write the expression of the probability distribution of user $u$'s preference for item $j$ given the user $a$'s preference for it as follows:

$Pr(S_{uj} = x_{uj} | S_{aj} = x_{aj}) = P(y = x_{uj} | x = x_{aj})$

$$= \frac{1}{\sqrt{2\pi\sigma^2}} exp\left[\frac{-1}{2\sigma^2}(x_{uj} - (\alpha + \beta x_{aj}))^2\right]$$

To estimate $Pr(S_{ak}|S_{aj})$, we use the same model as described above for computing $Pr(S_{uj}|S_{aj})$. In this case the independent variable $x$ denotes the rating received by item $j$, while dependent variable $y$ denotes the rating received by item $k$. And, the $n$ pairs of observations $(x_1, y_1), (x_2, y_2), \ldots, (x_n, y_n)$ are the ratings over item $j$ and $k$ by those n users, who have co-rated both items $j$ and $k$.

After computing the probability distribution $Pr(S_{uj}|S_{aj})$ for all users ($u$) who rated item $j$, and $Pr(S_{ak}|S_{aj})$ for all items ($k$) rated by user $a$, we can compute the probability distribution of the user $a$ rating for item $j$ using Equation (1). In this model we need to compute $3*(u+i)$ free parameters ($\alpha$, $\beta$, and $\sigma^2$), where $u$ is the number of users who rated the item $j$ and $i$ is the number of items rated by user $a$.

When the linear relationship exceeds the maximum value of the rating scale, we use the maximum value; when it is lower than the minimum value of the rating scale, we use the minimum value.

To predict a rating (for example, to compare it with other algorithms that predict ratings), we predict the expected value of the rating. The expected value of the rating is defined as follows:

$E(S_{aj}) = \sum_{i=1}^{m} Pr(S_{aj} = v_i|X) * v_i$

where $X = (S_{1j}, \ldots, S_{Nj}) \wedge (S_{a1}, \ldots, S_{aM})$.

### 3.1  $K$ Dummy Observations

While trying to fit lines with sparse data, we often find a perfect linear relationship, even though the sensor isn't perfect. If there is a perfect fit between users $a$ and $u$, then the variance will be zero according to the above calculations. Therefore, the sensor $u$'s prediction for item $j$ will be perfect, or deterministic; that is, the conditional probability table associated with sensor $u$ will be purely deterministic. We do not want this for our noisy sensor model because a deterministic sensor will discount the effects of other sensors. For example, often one or two co-rated items always have a perfect fit, even though such a user is not a good sensor.

We hypothesize that this problem can be overcome if we add $K$ dummy observations along with $n$ observations (co-rated items). We assume that user $a$ and user $u$ give ratings over $K$ dummy items ($K > 0$) in such a way that their ratings for $K$ dummy items are distributed over all possible rating pairs. For $m$ scale rating data there are $m^2$ possible combination of the rating pairs. We compute the prior distribution of each rating pair by its frequency in the training data. We use the prior distribution of rating pairs for distributing the effect of $K$ dummy points over all rating pairs like *hierarchical prior*. This, however, reduces our ability to guarantee the ratings for $K$ items will be distributed over all possible rating pairs. We have experimented with parameter $K$, and we found that *Noisy1* performs better with $K = 1$. For subsequent experiments we, therefore, chose $K = 1$ for *Noisy1*.

### 3.2  Selecting Noisy Sensors

For determining the reliability of the noisy sensors, we consider the *goodness of fit* of the fitted regression line to a set of observations. We use the coefficient of determination $r^2$ [Gujarati, 1995], a measure that tells how well the regression line fits the observations. This coefficient measures the proportion or percentage of the total variation in the dependent variable explained by the regression model. $r^2$ is calculated as follows [Gujarati, 1995]:

$$r^2 = 1 - \frac{\sum e_i^2}{\sum(y_i - \overline{y})^2}$$

where $\overline{y}$ is the mean of the ratings of user $u$.

The value of $r^2$ lies between 0 and 1 ($0 \leq r^2 \leq 1$). If $r^2 = 1$, there exists a perfect linear relationship between the preferences of user $a$ and user $u$; that is, $e_i = 0$ for each observation (co-rated item). On the other hand, if $r^2 = 0$, it means there is no linear relationship between users $a$ and $u$ and the best fit line is horizontal line going through the mean $\overline{y}$. We order the user and item noisy sensors according to $r^2$. We use the best $U$ user noisy sensors and best $I$ item noisy sensors for making the predictions. The parameter settings for $U$ and $I$ are explained in the next section.

### 3.3  Variant *Noisy2* of *Noisy1*

The problem with *Noisy1* is that we must often fit linear relationships with very little data (co-rated items). It may be better to assume *a priori* the linear model and then simply learn the noise. The algorithm *Noisy2* is based on the idea that different users rate items the same and, similarly, different items receive the same rating. We assume that the preferences of user $a$ and user $u$ are the same; that is, the expected value of user $u$'s preference of any item is equal to active user $a$'s preference for that item.

$$E(y_i|x = x_i) = \mu = x_i$$

We learn the variance of user $u$'s prediction. The algorithm *Noisy2* can be derived from algorithm *Noisy1*

by making the regression coefficients $\alpha = 0$ and $\beta = 1$. In this model we need to compute $(u + i)$ free parameters ($\sigma^2$), where $u$ is the number of users who rated the item $j$ and $i$ is the number of items rated by user $a$. We also add the $K$ dummy observations because the same problem (as discussed in Subsection 3.1) can arise in *Noisy2*. We have experimented with parameter $K$, and we found that *Noisy2* also performs better with $K = 1$. For subsequent experiments we, therefore, chose $K = 1$ for *Noisy2* also.

In *Noisy2* we are not fitting the relationship between user $a$ and user $u$, but we assume an equal relationship. So, it doesn't make sense to use the coefficient of determination $r^2$ for finding the reliability of the noisy sensors. Rather, to find the reliability of the noisy sensor, we use the variance; the less variance, the more reliable the noisy sensor. We use the best $U$ user noisy sensors and best $I$ item noisy sensors for making the predictions. The parameter settings for $U$ and $I$ are explained in the next section.

## 4 Evaluation Framework

To evaluate the accuracy of collaborative filtering algorithms we used the *training and test set* approach. In this approach, the dataset of users (and their ratings) is divided into two: a *training set* and a *test set*. The *training set* is used as the collaborative filtering dataset. The *test set* is used to evaluate the accuracy of the collaborative filtering algorithm. We treat each user from the test set as the *active* user. To carry out testing, we divide the ratings by each test user into two sets: $I_a$ and $P_a$. The set $I_a$ contains ratings that we treat as observed ratings. The set $P_a$ contains the ratings that we attempt to predict using a CF algorithm and observed ratings ($I_a$) and training set.

To evaluate the accuracy of the collaborative filtering algorithm we use the average absolute deviation metric, as it is the most commonly used metric [Breese et al., 1998]. The lower the average absolute deviation, the more accurate the collaborative filtering algorithm is. For all users in the test set we calculate the average absolute deviation of the predicted rating against the actual rating of items. Let the number of predicted ratings in the test set for the active user be $n_a$; then the average absolute deviation for a user is given as follows:

$S_a = \frac{1}{n_a} \sum_{j \in P_a} |p_{a,j} - r_{a,j}|$,

where $p_{aj}$ is user $a$'s observed rating for item $j$ and $r_{aj}$ is user $a$'s predicted rating for item $j$.

These absolute deviations are then averaged over all users in the test set.

### 4.1 Data and protocols

We compared both versions of our noisy sensor model to *PD* (Personality Diagnosis) [Pennock et al., 2000] and *Correlation* (Pearson Correlation) [Resnick et al., 1994]. To compare the performance we used the same subset of the EachMovie database as used by Breese et al. [Breese et al., 1998] and Pennock et al. [Pennock et al., 2000], consisting of 1,623 movies, 381,862 ratings, 5,000 users in the training set, and 4,119 users in the test set. In EachMovie database the ratings are elicited on a integer scale from zero to five. We also tested the algorithms on other subsets to verify that we are not finding peculiarities of the subset.

We ran experiments with different amounts of ratings in set $I_a$ to understand the effect of the amount of the observed ratings on the prediction accuracy of the collaborative filtering algorithm. As in [Breese et al., 1998] for the *AllBut1 Protocol*, we put a single randomly selected rating for each test user in the test set $P_a$ and the rest of the ratings in the observed set $I_a$. As in [Breese et al., 1998] for each *GivenX Protocol*, we place $X$ ratings randomly for each test user in the observed set $I_a$, and the rest of the ratings in the test set $P_a$. We did the experiments for $X = 2, 5$ and 10.

### 4.2 Selecting Noisy Sensors

After learning the noisy sensor model we determine which noisy sensors should be used in making the prediction for the active user. Figure 2 shows the variation of average absolute deviation with *best* user noisy sensors for different numbers of *best* item noisy sensors for *Noisy1*. We used the dataset as described above but the test rating and the observed ratings for each user of the test set were selected randomly.

Figure 2 shows that the average absolute deviation decreases with the increase in number of item sensors. There is no significant improvement in accuracy when the number of item sensors is more than twenty. It also shows that the average absolute deviation first decreases with the increase in number of user sensors and then increases as more user sensors are used for prediction. This is because the large number of user sensors results in too much noise for those user sensors that have good reliability.

From the experiments, we found that both algorithms give better performances with ten-to-twenty item noisy sensors and forty-to-seventy user noisy sensors [2]. For the experiments reported in the following section, we use the best fifty user noisy sensors and best twenty item noisy sensors (i.e. $U = 50$ and $I = 20$) for both *Noisy1* and *Noisy2*. The parameters $U$ and $I$ depend on the database. In the case of the EachMovie database, the number of users are more than the movies, also each user has rated only few movies. Due to this possibility more best user noisy sensors ($U$) are selected than the best item noisy sensors ($I$).

---

[2] We didn't use the test set for finding the number of user and item noisy sensors.

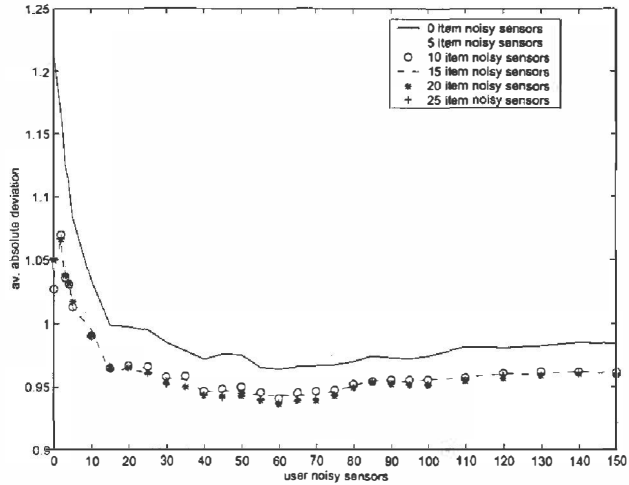

Figure 2: The average absolute deviation as a function of the number of best user noisy sensors for different numbers of best item noisy sensors.

From Figure 2 we also see that the minimum average absolute deviation is .936 when we use both user and item noisy sensors (with sixty user and twenty item noisy sensors). It is .964 when we use only the user noisy sensors, shown by the zero item noisy sensors case, and 1.027 when we use only the item noisy sensors, shown on the y-axis for ten item noisy sensors. This indicates that when we include the item noisy sensors along with the user noisy sensors, the quality of the prediction improves considerably. It also shows that if we use only the item noisy sensors for prediction, then the average absolute deviation becomes greater than when we use only user noisy sensors. Therefore, symmetric collaborative filtering offers better accuracy than asymmetric collaborative filtering.

### 4.3 Comparison with Other Methods

We compared the algorithms *Noisy1*, *Noisy2*, *Correlation* and *PD* using the same training and test set as Pennock et al. [Pennock et al., 2000]. For each test user in the test set we use the same set of observed ($I_a$) and test ratings ($P_a$) as Pennock et al.

The results of comparing *Noisy1*, *Noisy2*, *Correlation* and *PD* are shown in Table 1. We re-implemented *Personality Diagnosis*. Our results for *Personality Diagnosis* match exactly with those reported in Pennock et al. [Pennock et al., 2000]. We took the results for *Correlation* directly from Pennock et al. [Pennock et al., 2000].

*Noisy1* performed better than *PD* and *Correlation* for *AllBut1* and *Given10* protocols. For *Given5* and *Given2* protocols *Noisy1* performance is better than *Correlation* but not better than *PD*. We believe that *Noisy1*'s poor performance can be explained by the fact the lines that are fitted to very small data sets are often poor fit to the actual relationship. The algorithm *Noisy2*, based on an equal relationship between users, doesn't suffer from the same problem, and outperformed all algorithms under all protocols.

Table 1: Average absolute deviation scores on the EachMovie data for *Noisy1*, *Noisy2*, *PD* and *Correlation* (note: lower scores are better).

| Algorithm | Protocol | | | |
|---|---|---|---|---|
| | AllBut1 | Given10 | Given5 | Given2 |
| Noisy2 | **.893** | **.943** | **.974** | **1.012** |
| Noisy1 | .943 | .983 | 1.021 | 1.196 |
| PD | .964 | .986 | 1.016 | 1.039 |
| Correl | .999 | 1.069 | 1.145 | 1.296 |

Shardanand and Maes [Shardanand and Maes, 1995] and Pennock et al. [Pennock et al., 2000] proposed that the accuracy of a collaborative filtering algorithm should be evaluated on extreme ratings (very high or very low ratings) for items. The supposition is that, most of the time, people are interested in suggestions about items they might like or dislike, but not about items they are unsure of. Pennock et al. [Pennock et al., 2000] defined the extreme ratings as those which are 0.5 above and 0.5 below the average rating in the subset. To compare the performance of algorithms with extreme ratings we computed the predicted ratings for those test cases from the test set $P_a$ of all protocols, whose observed rating is less than $\overline{R} - 0.5$ or greater than $\overline{R} + 0.5$, where $\overline{R}$ is the overall average rating in the subset.

Table 2: Average absolute deviation scores on EachMovie data for *Noisy1*, *Noisy2*, *PD* and *Correlation* for extreme ratings.

| Algorithm | Protocol | | | |
|---|---|---|---|---|
| | AllBut1 | Given10 | Given5 | Given2 |
| Noisy2 | 1.001 | **1.057** | **1.087** | **1.124** |
| Noisy1 | **.997** | 1.063 | 1.125 | 1.249 |
| PD | 1.029 | 1.087 | 1.128 | 1.163 |
| Correl | 1.108 | 1.127 | 1.167 | 1.189 |

The results for the extreme ratings are shown in Table 2. The results for extreme ratings show that *Noisy1* performs better than *Noisy2* for *AllBut1* protocol. It also performs better than *PD* and *Correlation* over *Given10* and *Given5* protocols. *Noisy2* performs better than the other three algorithms over *Given10*, *Given5* and *Given2* protocols.

Table 3: Significance levels of the differences in average absolute deviation between *Noisy1* and *PD*, and between *Noisy2* and *PD*, on EachMovie data (note: low significance levels indicate that the differences in results are unlikely to be coincidental).

| Protocol | *Noisy1* vs. *PD* | *PD* vs. *Noisy1* | *Noisy2* vs. *PD* | *PD* vs. *Noisy2* |
|---|---|---|---|---|
| *AllBut1* | .0006 | .9994 | .0001 | .9999 |
| *AllBut1* (extreme) | .0003 | .9997 | .0127 | .9873 |
| *Given10* | .1377 | .8623 | .0001 | .9999 |
| *Given10* (extreme) | .0211 | .9789 | .0009 | .9991 |
| *Given5* | .9064 | .0936 | .0043 | .9957 |
| *Given5* (extreme) | .2897 | .7103 | .0001 | .9999 |
| *Given2* | .9999 | .0001 | .0019 | .9981 |
| *Given2* (extreme) | .9999 | .0001 | .0001 | .9999 |

To determine the statistical significance of these results, we computed the *significance levels* for the differences in average absolute deviation between *Noisy1* and *PD*, *PD* and *Noisy1*, *Noisy2* and *PD*, and *PD* and *Noisy2* for all protocols. To do this, we divided the test set for all protocols into 60 samples of equal size and used randomization paired sample test of differences of means [Cohen, 1995]. This method calculates the sampling distribution of the mean difference between two algorithms by repeatedly shuffling and recalculating the mean difference in 10,000 different permutations. The shuffling reverses the sign of the difference score for each sample with a probability of .5.

The statistical significance results of the EachMovie data results are shown in Table 3; it shows the probability of achieving a difference less than or equal to the mean difference. That is, it shows the probability of incorrectly rejecting the null hypothesis that both algorithms' deviation scores arise from the same distribution.

## 5 Conclusion

In this paper, we have concerned ourselves with symmetric collaborative filtering based on explicit ratings used for making recommendations to a user based on ratings of various items by a number of people, and the user's ratings of various items.

We have described a new probabilistic approach for symmetric collaborative filtering based on a noisy sensor model. We have shown that the noisy sensor model makes better predictions than other state-of-the-art techniques. The results for *Noisy2* are highly statistically significant. We have also shown that by including the items similarity along with users similarity, the accuracy of prediction increases. This paper has only considered the accuracy of the noisy sensor model, not on the computational issues involved. It is beyond the scope of this paper to consider the trade-off between off-line and online computation and effective indexing to find the best matches.

## 6 Acknowledgments

We thank Compaq Equipment Corporation and David M. Pennock for providing the EachMovie database and subsets used in this study. We also thank Holger Hoos for providing the valuable comments. This work was supported by Institute for Robotics and Intelligent Systems project BOU and the Natural Sciences and Engineering Research Council of Canada Operating Grant OGP0044121.

## References

Billsus, D. and Pazzani, M. J. [1998]. Learning collaborative information filters, *Proceedings of the 15th International Conference on Machine Learning*, pp. 46–54.

Breese, J. S., Heckerman, D. and Kadie, C. [1998]. Empirical ananlysis of predictive algorithms for collaborative filtering., *Proceedings of the 14th Conference on Uncertainity in Artificial Intelligence*, pp. 43–52.

Cohen, P. R. [1995]. *Empirical Methods for Artificial Intelligence*, MIT Press.

Gujarati, D. N. [1995]. *Basic Econometrics*, third edn, McGraw Hill, Inc.

Pennock, D. M., Horvitz, E., Lawrence, S. and Giles, C. L. [2000]. Collaborative filtering by personality diagnosis: A hybrid memory-and model-based approach., *Proceedings of the 16th Conference on Uncertainty in Artificial intelligence*, pp. 473–480.

Resnick, P., Iacovou, N., Suchak, M., Bergstrom, P. and Riedl, J. [1994]. Grouplens: An open archi-

tecture for collaborative filtering of netnews, *Proceedings of ACM CSCW'94 Conference on Computer Supported Cooperative Work*, pp. 175–186.

Schafer, J. B., Konstan, J. and Riedl, J. [1999]. Recommender system in e-commerce., *Proceedings of the ACM Conference on Electronic Commerce (EC-99)*, pp. 158–166.

Shardanand, U. and Maes, P. [1995]. Social information filtering: Algorithms for automating "word of mouth"., *Proceedings of ACM CHI'95 Conference on Human Factors in Computing Systems*, pp. 210–217.

Sharma, R. [2001]. *Symmetric Collaborative Filtering using the Noisy Sensor Model*, M.Sc. thesis, http://www.cs.ubc.ca/~rsharma.

Ungar, L. H. and Foster, D. P. [1998]. Clustering methods for collaborative filtering, *Workshop on Recommendation Systems at the 15th National Conference on Artificial Intelligence*.